\documentclass[10pt, conference]{IEEEtran}
\usepackage{amsmath,amssymb,amsthm,mathrsfs,amsfonts,dsfont,mathtools}
\usepackage[caption=false]{subfig}
\usepackage[ruled,lined,linesnumbered]{algorithm2e}
\usepackage{setspace}
\SetAlFnt{\small}
\SetAlCapFnt{\small}
\SetAlCapNameFnt{\small}

\usepackage{cite, color}
\usepackage{graphicx,epstopdf}
\usepackage{bmpsize}
\usepackage{hyperref}

\def\eps{\varepsilon}
\makeatletter
\newcommand{\vast}{\bBigg@{2.7}}
\newcommand{\Vast}{\bBigg@{4}}
\makeatother

\newtheorem{theorem}{Theorem}
\newtheorem{lemma}{Lemma}

\ifCLASSINFOpdf
\else
\fi

\ifCLASSOPTIONcaptionsoff
  \newpage
\fi

\begin{document}

\title{Delay and Backlog Analysis for $ 60 $ GHz Wireless  Networks}

\author{\IEEEauthorblockN{Guang Yang, Ming Xiao, James Gross, and Hussein Al-Zubaidy}
\IEEEauthorblockA{Communication Theory Department\\
Royal Institute of Technology,
Stockholm 100 44, Sweden\\
Email: \{gy,mingx,hzubaidy\}@kth.se, james.gross@ee.kth.se}
\and
\IEEEauthorblockN{Yongming~Huang}
\IEEEauthorblockA{School of Information Science
and Engineering\\
Southeast University, Nanjing 210096, China\\
Email: huangym@seu.edu.cn}
}

\maketitle

\begin{abstract}
To meet the ever-increasing demands on higher throughput and better  network delay performance, $ 60 $~GHZ networking is proposed as a promising   solution for the next generation of wireless communications. 
To successfully deploy such networks, its important to understand their performance first. However, due to the unique  fading characteristic of the $ 60 $~GHz channel, the characterization of the corresponding service process, offered by the channel,  using the conventional methodologies may not be tractable.
In this work, we provide an alternative approach to derive a closed-form expression that characterizes  the cumulative service process of the $ 60 $~GHz channel in terms of the moment generating function (MGF) of its instantaneous channel capacity.
We then use this expression to derive probabilistic upper bounds on the backlog and delay that are experienced by a flow traversing this network, using results from the MGF-based network calculus.
The computed bounds are validated using simulation. We provide numerical results for different networking scenarios and for different traffic and channel parameters and we show that the 60 GHz  wireless network is capable of satisfying stringent quality-of-Service (QoS) requirements, in terms of network delay and reliability. With this analysis approach at hand, a larger scale 60 GHz network design and optimization is  possible.

\end{abstract}

\begin{IEEEkeywords}
$ 60 $ GHz; Moment Generating Functions; Backlog; Delay; Upper Bound.
\end{IEEEkeywords}

\section{Introduction}\label{sec:Intro}
In order to satisfy the rapidly increasing demands on higher network performance, e.g., in terms of higher quality-of-experience (QoE) \cite{pierucci2015quality}, quality-of-service (QoS) \cite{hossain2014evolution}, and greater mobile data traffic \cite{cisco2015visual}, wireless communications using $ 60 $ GHz radio (often referred to as millimeter wave: mmWave communication system) is proposed as a promising alternative to the existing lower band (around 2 GHz) communication. To date, there have been several academic and industrial bodies participating in the standardization of mmWave technology, e.g., IEEE 802.11ad Task Group, IEEE 802.15.3 Task Group 3c, and Wireless Gigabit Alliance (WiGig). There are also numerous  efforts in academia that are  dedicated to   network design and related applications for the next generation wireless communications. However, due to the  severe path loss and high oxygen absorption effect in the $ 60 $ GHz radios \cite{daniels200760,geng2009millimeter}, it is not yet clear if such networks can meet the desired delay and reliability requirements. 
Furthermore, performance evaluations for the $ 60 $ GHz wireless networks are of great importance in the system design, optimization and implementation, if the performance requirements, by the end users, are to be guaranteed. Unfortunately, the traditional performance analysis methodologies, e.g., classical queuing theory, may not be useful in analysing such networks unless some conformity assumptions on the arrival and service process are enforced. Such assumptions may jeopardize the integrity of the obtained results in reference to the real system.  To date, most of related study regarding the performance of $ 60 $ GHz wireless networks departs from the perspective of physical layer (PHY), where a series of PHY techniques, e.g., signal modulation, channel coding and beamforming, were discussed \cite{choi2009performance, rakotondrainibe2010performance}. However, a network layer delay and backlog analysis of $ 60 $ GHz wireless communications in terms of the channel parameters, which is crucial in evaluating the performance of QoE- or QoS-driven wireless communications, does not exist. This motivates us to develop an analytical model and methodology to analyze the performance of such networks.

We propose an alternative and more suitable   methodology to analyze the performance of 60 GHz networks  based on a network calculus approach. This approach was  originally proposed by Cruz \cite{cruz1991calculus} for worst-case analysis of deterministic networked systems. Network calculus is a system theoretic approach for the analysis of communication networks, where the network operation is described  using the $ (\min,+) $~algebra. A network service element, e.g., a wired or wireless link,  is modelled as a $ (\min,+) $~linear, time-invariant (LTI) system and the input/output relationship is governed by the LTI system theory in the $ (\min,+) $~algebra. Meanwhile, the original theory for deterministic systems analysis has been   extended  to  probabilistic settings in order to model and analyze stochastic networks \cite{jiang2008stochastic}. One approach to extend network calculus to stochastic settings is the MGF-based network calculus that was proposed by \cite{fidler2010survey} which extends initial results regarding the MGF arrival and service curves suggested  in \cite{chang2000performance}. More recently, a wireless network calculus based on the $ (\min,\times) $~algebra was proposed \cite{alnetwork}. In this approach, in order  to simplify an otherwise intractable analysis, the network model is transferred into an alternative analysis domain (referred to as SNR domain) by using the   exponential function.  Probabilistic end-to-end performance bounds are provided for homogeneous multi-hop channels and numerical results for the  case of multi-hop Rayleigh fading channels is presented.

The MGF-based network calculus \cite{fidler2010survey} was originally developed to enable the modelling of the   multiplexing gain of many independent stochastic flows. It was later used in the  analysis of  wireless systems by characterising the system's service process in terms of the MGF of a Markov channel model of the wireless link \cite{fidler2006wlc15,mahmood2011flow}.

Although network calculus has been around for some years, its applications to wireless networks are fairly recent, and the investigated scenarios are limited to Rayleigh fading channels. In contrast to the existing work on analyzing the performance of wireless networks, The $ 60 $ GHz radio channel present several challenges. First, unlike conventional wireless communications in the bands below $5$ GHz, which experience small-scale fading effects \cite{williamson1997investigating}, the $ 60 $ GHz fading channels  mainly experience log-normally distributed shadowing fluctuations \cite{rangan2014millimeter, rappaport2015wideband}. Second, due to the particular fading characteristic of $ 60 $ GHz channels, the  $ \left(\min, \times\right) $~based wireless network calculus presented in \cite{alnetwork} is not applicable to the analysis of this network\footnote{The Mellin transform for the log-normal distribution does not exist, therefore the computation of performance bounds by $ \left(\min,\times\right) $ algebra in 'SNR domain', which is transferred from the 'bit domain', is not feasible.}. Therefore, we opt to use the MGF-based network calculus instead. Finally, the service process characterization of the 60 GHz channel, i.e., computing the MGF of the channel capacity, is generally intractable which makes it difficult to obtain a closed-form exact solution. We use  clever manipulations and bounding of the resulted expression to obtain a tractable bound on the desired service process. 

In this work, we provide analytical  probabilistic backlog and delay bounds for the $ 60 $ GHz wireless access network. This in turn corresponds to networked applications with loss- and latency-sensitivity respectively. To do that,  we first derive a simplified closed-form expression for the network service curve, in terms of the  fading model and parameters of the $ 60 $ GHz channel. We then evaluate the performance of the proposed model using the MGF-based network calculus approach and derive the desired probabilistic  bounds. Finally, we discuss the impact of a series of factors, such as arrival rate, system gain and channel fading characteristic, on network performance. 
Our novel application of MGF-based network calculus to the analysis of   60 GHz wireless networks   reveals useful insights for the network operation and the effect of different channel parameters on network performance. It also demonstrates the ability of such networks to satisfy the stringent service requirements of  modern networked applications. 

The remainder of the paper is organized as follows. Basics for MGF-based stochastic network calculus are given in Sec.~\ref{Sec: II}. Probabilistic backlog and delay bounds for the $ 60 $ GHz wireless network are derived in Sec.~\ref{Sec: III}. Numerical results are presented in Sec.~\ref{Sec: V}. Finally, conclusions are drawn in Sec.~\ref{Sec: VI}.

\section{MGF-based Network Calculus}\label{Sec: II}

\begin{figure}
\centering
\label{fig: queuing model}
\includegraphics[width=2.5in]{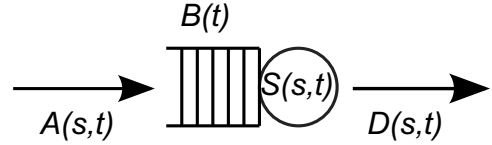}
\caption{A  queuing model for a store-and-forward node.}
\end{figure}

Assuming a fluid-flow, discrete-time queuing system with a buffer of infinite size, and given a time interval $[s,t)$, $0\le s \leq t $, we define the  non-decreasing (in $t$) bivariate processes $A(s,t), D(s,t)$ and $S(s,t)$ as the cumulative arrival to, departure from and service offered by the system as illustrated in Fig.~\ref{fig: queuing model}. We further assume that $A,D$ and $S$ are stationary non-negative random processes with $A(t,t)= D(t,t)= S(t,t)=0$ for all $t\ge 0$. In addition, throughout the paper we only consider causal systems, i.e., $ D(s,t)\leq A(s,t) $. We denote by $ B(t) $  the backlog (the amount of buffered data) at time $ t $. More details and the proofs for the following presented fundamental results can be found in \cite{le2001network,chang2000performance}.

\subsection{Traffics and Service Characterizations}\label{sec: II_A}
In network calculus, given the service process $ S(s,t) $, the departure process $D(s,t)$ is related to the arrival process $A(s,t)$ via the $ (\min,+) $ convolution denoted by ($\otimes$). Another important operation in the $(\min,+)$ system theory is the $ (\min,+) $, deconvolution denoted by  ($\oslash$). Here, $ \otimes $ and $ \oslash $ are respectively defined as
\begin{equation*}
(X \otimes Y)(s,t) \triangleq \inf_{s\leq \tau \leq t}\{X(s,\tau)+Y(\tau,t) \},
\label{eqn:IIa1}
\end{equation*}
\begin{equation*}
(X \oslash Y)(s,t) \triangleq \sup_{0\leq\tau\leq s}\{X(\tau,t)-Y(\tau,s) \},
\label{eqn:IIa2}
\end{equation*}
where $ X $ and $ Y $ are arbitrary bivariate functions. For time varying systems, we refer to a network element as \textit{dynamic server}  if for all $ t\geq 0 $ it offers a time varying service $S$ that satisfies the following input-output inequality  \cite{chang2000performance}
\begin{equation}\label{eqn:dynamic server}
D(0,t)\geq (A\otimes S)(0,t)\,.
\end{equation}
The above holds with strict equality when the system is linear \cite{le2001network}. One typical example of \eqref{eqn:dynamic server} is a work-conserving link with a time-variant capacity, with the available service $ S(s,t) $ during interval $ [s,t) $.

An important and frequently used traffic envelope that considers traffic burstiness as well as the sustainable arrival rate is the affine envelope \cite{le2001network}, i.e., $A(s,t) = \rho(t-s)+\sigma$. This envelope was later extended to probabilistic settings as the $(\sigma(\theta),\rho(\theta))$ traffic envelope  \cite{chang2000performance}, in the sense that it provides, for a given $\theta$, a linear bound on the $\log$ MGF of the arrival process, i.e., 
\begin{equation}\label{eq:arrival-model} 
\frac{1}{\theta} \ln \mathbb{E}\left[e^{\theta A(s,t)}\right ] \le \rho(\theta)\cdot (t-s)+ \sigma(\theta).
\end{equation} 

The cumulative arrival and service  processes  are given in terms of their instantaneous values, $ a_{i} $ and $ s_{i} $ respectively during the $ i^{\mathrm{th}} $ time slot, as $ A(\tau,t)=\sum_{i=\tau}^{t-1}a_{i} $ and $ S(\tau,t)=\sum_{i=\tau}^{t-1}s_{i} $
for all $0 \le \tau \le t$.
For wireless networks,  $ s_{i} $ represents the instantaneous fading channel capacity during the $ i^{\mathrm{th}} $ time slot, which is given by $ s_i = W \ln (1+ \gamma(i)) $
`nats' per second, where $W$ is the channel bandwidth and $\gamma(i)$ is the instantaneous  SNR at the receiver during time slot $i$.

\subsection{Backlog and Delay Bounds}\label{sec: II_B}
For a given queuing system with cumulative arrival $ A(0,t) $ and departure $ D(0,t) $ and for $t\ge 0$, the backlog $ B\left(t\right) $ and virtual delay $W(t)$ in a first-come-first-serve (FCFS) scheduling system are respectively given by
\begin{equation}\label{eqn:backlog1}
B\left(t\right) \triangleq A\left(0,t\right)-D\left(0,t\right)\leq \left(A\oslash S\right)\left(t,t\right),
\end{equation}
and
\begin{equation}\label{eqn:delay1}
\begin{split}
W(t)\triangleq & \inf\{w\geq 0: A(0,t)\leq D(0,t+w) \} \\
 \leq & \inf\{w\geq 0: (A\oslash S)(t+w,t)\leq 0 \},
\end{split}
\end{equation}
where the inequalities are obtained by substituting  \eqref{eqn:dynamic server} in the above expressions and then using the definitions of $ \otimes $ and $ \oslash $.


In  deterministic network calculus, only the worst-case upper bound is considered \cite{le2001network} which result in pessimistic performance bounds that does not capture the multiplexing gain effect in packet networks. Furthermore, deterministic analysis  cannot  be used for wireless network performance analysis since $S(\tau,t)=0$ is the only possible deterministic lower service curve due to the possibility of outage in wireless networks. 
The stochastic analysis on the other hand, can surpass the above limitations. In the probabilistic setting, the backlog and delay bounds with respect to \eqref{eqn:backlog1} and \eqref{eqn:delay1} are  expressed as  
\begin{equation}\label{eqn:probabilistic bounds}
\mathbb{P}\left(B(t)>b^{\eps'}\right)\leq \eps' \; \mbox{and} \; \mathbb{P}\left(W(t)>w^{\eps''}\right)\leq \eps'',
\end{equation}
respectively, where $ b^{\eps'} $ and $ w^{\eps''} $  denote the probabilistic backlog and delay bounds, accordingly associated with $ \eps' $ and $ \eps'' $ that are defined as \emph{violation probabilities}. The probabilistic bound is usually computed using the distributions of these processes, i.e., in terms of the arrival and service processes MGFs \cite{fidler2010survey} or their Mellin transforms \cite{alnetwork}. In general, the MGF-based bounds are obtained by applying the Chernoff's bound, that is, given a random variable $X$, it is known that $ \mathbb{P}\left(X \geq x\right)\leq e^{-\theta x} \mathbb{E}\left[e^{\theta X}\right]\triangleq e^{-\theta x}\mathbb{M}_{X}(\theta) $,
whenever the expectation exists, where $ \mathbb{E}\left[Y \right] $ and $ \mathbb{M}_{Y}\left(\theta\right) $ denote the expectation  and the moment generating function (or the Laplace transform) of $ Y $, respectively, and  $ \theta $ is an arbitrary non-negative free parameter.
For the bivariate stationary processes $ X(s,t), t\ge s $, we define its MGF  \cite{fidler2006end} as 
\begin{equation*}
\mathbb{M}_{X}(\theta,s,t)\triangleq\mathbb{E}\left[e^{\theta X(s,t)}\right]\, .
\end{equation*}
Likewise, $ \overline{\mathbb{M}}_{X}(\theta,s,t)\triangleq \mathbb{M}_{X}(-\theta,s,t)= \mathbb{E}\left[e^{-\theta X(s,t)}\right] $ is also defined. 
Then, probabilistic backlog and delay bounds satisfying \eqref{eqn:probabilistic bounds}, can be respectively given by \cite{fidler2006end}
\begin{equation}
b^{\eps'} = \inf_{\theta >0}\left\lbrace \frac{1}{\theta} \left(\log \mathsf{M}\left(\theta,t,t\right)-\log \eps'\right) \right\rbrace,
\label{eqn: backlog bound}
\end{equation}
\begin{equation}
w^{\eps''}=\inf\left\lbrace w : \inf_{\theta>0} \left\lbrace \mathsf{M}(\theta,t+w,t) \right\rbrace \leq \eps'' \right\rbrace,
\label{eqn: delay bound}
\end{equation}
where $ \mathsf{M}(\theta,s,t) $ is shown as  
\begin{equation}
\mathsf{M}(\theta,s,t)\triangleq\sum_{u=0}^{\min(s,t)}\mathbb{M}_{A}(\theta,u,t)\cdot \overline{\mathbb{M}}_{S}(\theta,u,s).
\label{eqn: generic MGF for backlog and delay}
\end{equation}

It is worth noting that, $ \mathsf{M}(\theta,s,t) $ is only valid when the arrival and service curves are independent. 

\section{Performance  of 60 GHz Wireless  Networks}\label{Sec: III}
In this section, we derive probabilistic backlog and delay bounds for the 60 GHz wireless network in Fig.~\ref{fig: network model}. We first obtain a service process characterization for the 60 GHz channel
based on a well accepted fading model in the literature, then apply the MGF-based network calculus to obtain the desired bounds.

The  $ 60 $ GHz communication system can be used as an access network to provide high rate data communication for the end users. Although such an application is theoretically possible, the user mobility and the line-of-sight requirement for the 60 GHz communication makes it impractical for now.  
 Alternatively, it can be used as part of  a multi-hop backhaul network whenever the wired alternatives, e.g., fibre optics, are too expensive or are not possible, perhaps due to difficult terrains and logistical issues.
 In such networks, the system in Fig.~\ref{fig: network model} may either be part of a 60 GHz multi-hop wireless backhaul network  or it is part of other type of high speed backhaul network such as  optical networks, as shown in Fig.~\ref{fig: queuing model in real world}, where high-rate transmissions are supported. The extension of our current  methodology to analyse multi-hop homogeneous/heterogeneous networks  is possible, but more involved, and thus, it is left for future work in order to conform with the page limit requirement.
 
\subsection{System Model}

\begin{figure}
\centering
\subfloat[A buffered wireless link.]
{\label{fig: network model}
\includegraphics[width=2.3in]{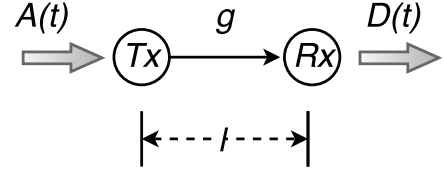}}

\subfloat[An illustration for acutal buffered wireless transmission.]
{\label{fig: queuing model in real world}
\includegraphics[width=2.7in]{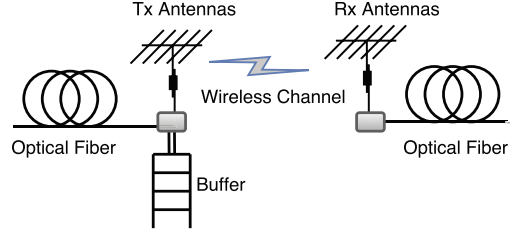}}

\caption{The analytical model and a corresponding real-world application of buffered $ 60 $~GHz wireless system.}
\end{figure}

In Fig.~\ref{fig: network model}, we denote by $ g $ and $ l $ the channel gain and separation distance, respectively. We adopt the system level assumptions stated in Sec. \ref{Sec: II} for he remainder of this section.
The small-scale fading effect in the $ 60 $ GHz communication system is negligible  \cite{geng2009millimeter}. This is due to the fact that the signal wavelength, denoted by $ \lambda $,  in the $ 60 $ GHz spectrum is very short, i.e., $ \lambda =5$ mm, which means that non-line-of-sight, i.e., reflected, components will most likely be absorbed and  attenuated quickly before they reach the destination.  Many research efforts have been devoted to characterize the indoor $ 60 $ GHz wireless personal area networks (WPAN) and outdoor measurements \cite{smulders2009statistical, rangan2014millimeter, yang2015maximum, rappaport2015wideband}, where a large amount of experimental results regarding this channel have been reported. In the light of above, a general expression for the channel gain (in dB), for either indoor or outdoor scenarios,   can be summarized as
\begin{equation}\label{eqn:channel gain}
g [\mathrm{dB}]=-\left(\alpha + 10\beta \log_{10} (l)+\xi\right),
\end{equation}
where, the separation distance $ l $ is given in meters, $ \alpha $ and $ \beta $ are the least square fits of floating intercept and slope of the best fit, and $ \xi\sim \mathcal{N}(0,\sigma^2) $ corresponds to the log-normal shadowing effect with variance $ \sigma^2 $. 

Using  Friis free space transmission formula   and the channel gain in \eqref{eqn:channel gain}, the signal-to-noise ratio (SNR), denoted by $ \gamma(k) $ during time slot $k$, can be expressed as
\begin{equation}\label{eqn:SNR}
\gamma(k) = \frac{P_{t}\cdot G_{Tx}\cdot G_{Rx}\cdot g_k}{N_{0}} \triangleq \kappa \cdot 10^{-\frac{\xi_k}{10}},
\end{equation}
where the constant prefactor $ \kappa $ represents the deterministic component of the system gain, which is determined by network configurations, including the transmit power $ P_{t} $, background noise power $ N_{0} $,  antennas gains $ G_{Tx} $ and $ G_{Rx} $ and their separation. In this work, we assume that the process $ \{\xi\}_{k=0}^\infty$  is stationary and independent in  time. 

\subsection{An MGF Bound  for the Cumulative Service Process}
The performance bounds stated in \eqref{eqn:probabilistic bounds} require the computation of the MGF of arrival and service processes. In this work, we use the 
$ \left(\sigma(\theta),\rho(\theta)\right) $ traffic characterization, described in \eqref{eq:arrival-model},
%
%
with parameters $ \sigma(\theta)=\delta_b $ and $ \rho(\theta)=\rho_a $, i.e., we assume deterministically bounded arrival process. 
Then \eqref{eq:arrival-model} reduces to 
\begin{equation} 
\mathbb{E}\left[e^{\theta A(s,t)}\right]\triangleq
\mathsf{M}_{A}(\theta,s,t)\leq 
e^{\theta \delta_b}\left(p_{a}(\theta)\right)^{t-s},
\label{eqn: MGF Arrival Process}
\end{equation}
for any $\theta > 0 $, where $ p_{a}(\theta)=e^{\theta\rho_{a}} $. Therefore, Equation \eqref{eqn: MGF Arrival Process} provides an upper bound on the MGF of the arrival process.
%

To characterize the service process, consider the following: given a channel SNR, $ \gamma(k) $ during time slot $k$, the instantaneous channel capacity    can be expressed as $ C_k=\eta\ln (1+\gamma(k) ) $
in bits/s, where $ \eta=\frac{W}{\ln 2} $ with channel bandwidth $ W $, then
the  cumulative service process, $ S(s,t) $, is given by
\begin{equation}\label{eq:MGF-service}
S(s,t) = \sum_{k=s}^{t-1} C_k = \eta\sum_{k=s}^{t-1}\ln \left(1+\gamma(k)\right),
\end{equation}
where $ \gamma(k) $ is defined by \eqref{eqn:SNR}. 

An exact expression for the MGF of the cumulative service process in \eqref{eq:MGF-service} is intractable. Instead, we  use Theorem \ref{theorem:upper bound of M_S}  that provides an upper bound  on the MGF of $ S(s,t) $. We first provide the  following lemma, which is essential for deriving this bound, before proceeding to Theorem \ref{theorem:upper bound of M_S}. 
\begin{lemma}\label{lemma:inverse moment}
Let $ F_{X}(x) $ denote the cumulative distribution function (c.d.f.) of non-negative random variable $ X $, then for $ \delta >0 $ and $ \theta \geq 0 $, we have $ \mathbb{E}\left[(1+X)^{-\theta}\right]
\leq \mathcal{U}_{\delta,x}(\theta) $, where
\begin{equation*}
\mathcal{U}_{\delta,x}(\theta)
= \min_{u\geq 0} \left\lbrace \left(1+\delta N_{\delta}(u) \right)^{-\theta} + \sum_{k=1}^{ N_{\delta}(u)} a_{\theta,\delta}(k) F_{X}(k\delta) \right\rbrace \,,
\end{equation*}
$ N_{\delta}(u)=\lfloor \frac{u}{\delta}\rfloor $ and $ a_{\theta,\delta}(k)= \left(1+(k-1)\delta\right)^{-\theta}-\left(1+k\delta\right)^{-\theta}$.
\end{lemma}
\begin{IEEEproof}
First, we  briefly show that $ \mathbb{E}\left[(1+X)^{-\theta}\right] $ exists for  non-negative $ X $ and $ \theta $. It is evident that, since $ \theta \geq 0 $, the inequality $ 0 < \left(1+X\right)^{-\theta} \leq 1$ generally  holds for any non-negative random variable $ X $. Let $ X_{i} $ with $ i=1,2,\ldots $ be i.i.d. instances from the sample space of $ X $, then by the law of large numbers (LLN), we immediately have
\begin{equation*}
\begin{split}
& \mathbb{E}\left[\big|(1+X)^{-\theta}\big|\right] =\mathbb{E}\left[(1+X)^{-\theta}\right]\\
=& \lim\limits_{N\rightarrow \infty} \frac{1}{N}\sum_{i=1}^{N}(1+X_{i})^{-\theta} \leq \lim\limits_{N\rightarrow \infty} \frac{1}{N}\sum_{i=1}^{N} 1 = 1 < \infty,
\end{split}
\end{equation*}
which proves the existence of the expectation. 

Let $ h(x)\triangleq\left(1+x\right)^{-\theta} $. Then $h(x)$ is monotone and convex in $x$ for any $ \theta \geq 0 $, since $ h'(x) = -\theta\left(1+x\right)^{-(\theta +1)}\leq 0 $ and $ h''(x) = \theta(\theta+1)\left(1+x\right)^{-(\theta +2)}\geq 0 $. We use this property next to obtain a bound on the desired expectation. Let $ f_{X}(x) = \frac{d}{dx} F_{X}(x)$ be the probability density function for the random variable $X$, then by partitioning  the range of $X$ to $(0, u)$ and $(u, \infty)$ and then  discretizing the lower range,  we obtain 
\begin{equation*}
\begin{split}
& \mathbb{E}\left[(1+X)^{-\theta}\right] \triangleq \int_{0}^{\infty}(1+x)^{-\theta} f_X(x)dx\\
\overset{(a)}{=} & \left(\int_{0}^{\delta} + \int_{\delta}^{2\delta} + \cdots + \int_{\delta\left(\lfloor \frac{u}{\delta}\rfloor-1\right)}^{\delta\lfloor \frac{u}{\delta}\rfloor} \right)(1+x)^{-\theta} f_X(x)dx \\
& + \int_{\delta\lfloor \frac{u}{\delta}\rfloor}^{\infty}(1+x)^{-\theta} f_X(x)dx\\
\overset{(b)}{\leq} & \min_{u\geq 0} \vast\lbrace \sum_{k=1}^{\lfloor \frac{u}{\delta}\rfloor}\left(1+(k-1)\delta\right)^{-\theta}\int_{(k-1)\delta}^{k\delta}f_X(x)dx \\
& +  \left(1+\delta\left\lfloor \frac{u}{\delta}\right\rfloor\right)^{-\theta}\int_{\delta\left\lfloor \frac{u}{\delta}\right\rfloor}^{\infty} f_X(x)dx \vast\rbrace\\
\overset{(c)}{=}& \min_{u\geq 0} \vast \lbrace \sum_{k=1}^{\lfloor \frac{u}{\delta}\rfloor}\left(1+(k-1)\delta\right)^{-\theta}\left(F_X(k\delta)-F_X((k-1)\delta)\right)  \\
& + \left(1+\delta\left\lfloor \frac{u}{\delta}\right\rfloor\right)^{-\theta}\left(1-F_X\left(\delta\left\lfloor \frac{u}{\delta}\right\rfloor\right)\right) \vast\rbrace,
\end{split}
\end{equation*}
where $ (a) $ is a reformulation by partitioning the integral region, $ (b) $ is achieved by applying the monotonicity and convexity of $ h(x)$, and $ (c) $ is due to the definition of 
$F_X(x)$ and the lemma follows by rearranging and combining similar terms.
%
\end{IEEEproof}
It is evident that, the tightness of the bound obtained in Lemma \ref{lemma:inverse moment} depends on the discretization step size $ \delta $. Technically, smaller step size yields a tighter upper bound. On the other hand, smaller step size costs more, computation wise, i.e.., the complexity is inversely propositional to the step size. 

\begin{theorem}\label{theorem:upper bound of M_S}
Given $ S(s,t)=\eta\sum_{k=s}^{t-1}\ln \left(1+\gamma(k)\right) $ with independent positive $ \gamma(k) $ for all $k=s, s+1, \dots t-1$, an upper bound on $ \overline{\mathbb{M}}_{S}(\theta,s,t) $ is given by
\begin{equation*}
\overline{\mathbb{M}}_{S}(\theta,s,t) \leq \prod_{k=s}^{t-1}q(\theta,k), \\
\label{eqn: upper bound of M_S}
\end{equation*}
where $ q(\theta,k) = \mathcal{U}_{\delta,\gamma(k)}\left(\eta\theta\right)$. Furthermore, if for all $ k $,  $ \gamma(k) \overset{\ell}{=} \gamma $, i.e. identically distributed $\gamma(k)$,   then $ q(\theta,k)=q(\theta) $   and the above expression  reduces to $ \overline{\mathbb{M}}_{S}(\theta,s,t) \leq \left ( q(\theta) \right )^{t-s}$.
\end{theorem}


\begin{IEEEproof}
Starting from the definition of $ \overline{\mathbb{M}}_{S}(\theta,s,t) $ and using the independence assumption of $ \gamma(k) $ in $k$, we have
\begin{equation*}
\begin{split}
 \overline{\mathbb{M}}_{S}(\theta,s,t)=& \mathbb{E}\left[e^{-\theta \cdot S(s,t)}\right]
=  \mathbb{E}\left[\prod_{k=s}^{t-1} e^{-\theta \eta \ln (1+\gamma(k))}\right] \\
=& \prod_{k=s}^{t-1}\mathbb{E}\left[\left(1+\gamma(k)\right)^{-\theta \eta}\right].
\end{split}
\end{equation*}
where we used \eqref{eq:MGF-service} in the first line, then using the independence assumption in the second step. 

Applying Lemma \ref{lemma:inverse moment}, the first part of Theorem \ref{theorem:upper bound of M_S} immediately follows. For the second part it suffice to  note that the distribution of a random variable is uniquely defined by its MGF. Then the second part follows.
%
%
\end{IEEEproof}

\subsection{Probabilistic Backlog and Delay Bounds}

Recalling the SNR expression in \eqref{eqn:SNR}, we note that the randomness in \eqref{eqn:channel gain} is due to the shadowing component which is assumed to follow the log-normal distribution. Therefore,  $ \gamma\left(k\right)\left[\mathrm{dB}\right]\sim \mathcal{N}\left(\kappa_{\mathrm{dB}},\sigma^2\right) $, i.e.,  normally distributed random variable, with mean $ \kappa_\mathrm{dB} $ and standard deviation $ \sigma $. Its c.d.f. $ F_{\gamma\left(k\right)}(x) $  is given by
\begin{equation}\label{eqn:lognormal distribution}
F_{\gamma\left(k\right)}(x) = \frac{1}{2}\mathrm{erfc}\left(-\frac{\ln x - \iota \kappa_{\mathrm{dB}}}{\sqrt{2}\iota \sigma}\right),
\end{equation}
where $ \iota = \frac{\ln 10}{10} $ and the complementary error function $ \mathrm{erfc}\left(x \right) $ is given by $ \mathrm{erfc}\left(x\right)\triangleq\frac{2}{\sqrt{\pi}}\int_{x}^{\infty}\exp\left(-t^2\right)dt $, and the process $ \{\xi\}_{k=0}^\infty$  is stationary and independent in  time, we immediately have
$ q(\theta) = \mathcal{U}_{\delta,\gamma(k)}\left(\eta\theta\right)$ by applying Lemma~\ref{lemma:inverse moment} and substituting \eqref{eqn:lognormal distribution} for the SNR distribution function.




Generally, the bound is tighter at smaller $\theta$. As the value of $\theta$ grows larger, the tightness of the bound becomes more sensitive to the step size parameter $\delta$.
Thus, the selection of appropriate value for  $ \delta $ is subject to the range of interest of the parameter $ \theta $ and it provides a trade-off between the quality of the obtained bounds and  computation requirements. 

Based on the previous descriptions of the arrival and service processes, Theorem~\ref{theorem:probabilistic bound} below provides the desired probabilistic backlog and delay bounds for the i.i.d. $\gamma(k)$ case. 

\begin{theorem}\label{theorem:probabilistic bound}
Let the arrival  and service processes be respectively characterized by $ p_{a}\left(\theta\right) $, $ \delta_{b} $ and $ q\left(\theta\right) $ as described in \eqref{eqn: MGF Arrival Process} and in Theorem \ref{theorem:upper bound of M_S}. 
Then, 
\begin{equation}\label{eq:TH2_B}
b^{\eps'} = \inf_{\theta >0}\left\lbrace \delta_{b}- {1 \over\theta} \Big( \log \left(1-p_{a}\left(\theta\right)q\left(\theta\right)\right) - \log\eps'\Big) \right\rbrace
\end{equation}
and
\begin{equation}\label{eq:TH2_D}
w^{\eps''} = \inf\left\lbrace w : \inf_{\theta>0} \left\lbrace \frac{e^{\theta \delta_b} q^{w}\left(\theta\right)}{1-p_{a}\left(\theta\right)q\left(\theta\right)} \right\rbrace \leq \eps'' \right\rbrace,
\end{equation}
are backlog and delay bounds satisfying \eqref{eqn:probabilistic bounds}, whenever the stability condition $  p_a(\theta)q(\theta)<1 $ holds.
\end{theorem}
\begin{IEEEproof} 
Substituting $ \mathbb{M}_{A} $ and $ \mathbb{M}_{S} $ from \eqref{eqn: MGF Arrival Process} and Theorem.~\ref{theorem:upper bound of M_S},  respectively, in \eqref{eqn: generic MGF for backlog and delay}, we obtain
%
\begin{equation}\label{eqn:upper bound of M}
\begin{split}
\mathsf{M}\left(\theta,s,t\right)
{\leq} & e^{\theta \delta_b} \sum_{u=0}^{\min(s,t)}p_{a}^{t-u}\left(\theta\right)\cdot q^{s-u}\left(\theta\right)\\
\overset{(a)}{=} & e^{\theta \delta_b} p_{a}^{t-s}\sum_{u=\tau''}^{s}\left(p_{a}\left(\theta\right)q\left(\theta\right)\right)^{u} \\
\overset{(b)}{\leq} & e^{\theta \delta_b} p_{a}^{t-s}\sum_{u=\tau''}^{\infty}\left(p_{a}\left(\theta\right)q\left(\theta\right)\right)^{u} \\
= &\frac{e^{\theta \delta_b} p_{a}^{\tau'}\left(\theta\right)q^{\tau''}\left(\theta\right)}{1-p_{a}\left(\theta\right)q\left(\theta\right)},
\end{split}
\end{equation}
where $ \tau'\triangleq\max(t-s,0) $ and $ \tau''\triangleq\max(s-t,0)$. Here,  equality $ (a) $ is obtained by performing the change of variables $u = s-u$ and rearranging terms. In $ (b) $ we let $s \rightarrow \infty$. The last step is obtained by evaluating the geometric sum. The sum converges only when $  p_a(\theta)q(\theta)<1 $ is satisfied.

The desired backlog and delay bounds in \eqref{eqn:probabilistic bounds} are obtained by substituting \eqref{eqn:upper bound of M} with $ s=t $ in  \eqref{eqn: backlog bound} and with $ s=t+w $ \eqref{eqn: delay bound} to obtain \eqref{eq:TH2_B} and \eqref{eq:TH2_B}, respectively. 
\end{IEEEproof}

\section{Numerical Results and Discussion}\label{Sec: V}
In this section, we provide numerical results and simulations for selected system configurations. The objective is two-fold:
\begin{enumerate}
\item to verify the obtained performance bounds;
\item to investigate and discuss system performance trends in order to draw conclusions and insights regarding the 60 GHz network performance.
\end{enumerate}

In what follows, simulations are realized via MATLAB. We assume constant rate arrival with rate $ \rho_a $ and  burst size $ \delta_b=0 $. Time is slotted  with intervals of $ 1 $ second.
The rest of the system parameters are listed in Table \ref{tab: system parameter}. Throughout this section, the system gain is given as $ \kappa=25 $ dB,  which is immediately obtained by applying \eqref{eqn:SNR} and the system setting.


\begin{table}[t]
\small
\renewcommand{\arraystretch}{1.5}
\centering
\caption{System Parameters}
\begin{tabular}{|c|c|c|}
\hline
\textbf{Parameters} & \textbf{Symbol} & \textbf{Value} \\
\hline
\hline
Bandwidth & $ W $ & $ 500 $ MHz\\
\hline
Transmit Power & $P_{t}$ & $ 1 $ mW \\
\hline
Antenna Gain & $ G_{t} $ and $ G_{r} $ & $ 20 $ dB\\
\hline
Noise Power Density& $N_0/W$ & $ -114 $ dBm/MHz\\
\hline
Separation Distance & $ l $ & $ 100 $ m\\
\hline
Path Loss Intercept & $\alpha$ & $ 70 $ \\
\hline
Path Loss Slope & $\beta$ & $ 2.45 $\\
\hline
\end{tabular}\label{tab: system parameter}
\end{table}

\begin{figure}
\centering
\includegraphics[width=3.4in]{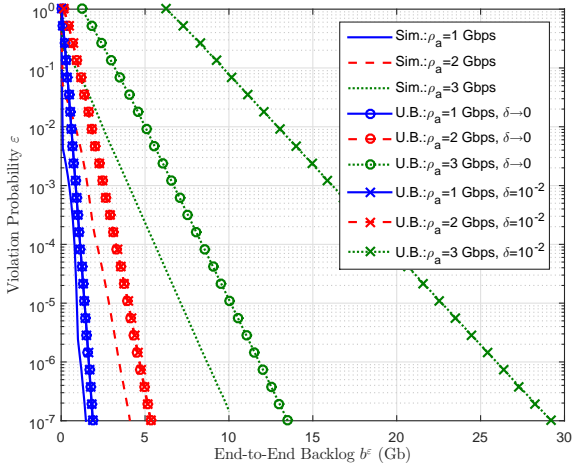}
\caption{Violation probability $ \eps $ v.s.  targeted backlog bounds $ b^{\eps} $, compared to simulation with $ \sigma=8 $ dB and $ \kappa=25 $ dB, for different $ \rho_a $.}
\label{fig:fig1_backlog}
\end{figure}

Fig.~\ref{fig:fig1_backlog} depicts the probabilistic  backlog bound compared to simulation. The arrival rate $ \rho_a $ is set to $ 1 $, $ 2 $ and $ 3 $ Gbps. We use two different values for the precision parameter, $ \delta=10^{-2} $ and $ \delta\rightarrow 0 $. As expected, the backlog increases as the arrival rate increases. The figure shows that the computed bounds capture the exponential decay of the backlog distribution. Furthermore, the bounds become asymptotically tighter, i.e., as $b^\eps \rightarrow \infty$. 

The derived result provided by Lemma \ref{lemma:inverse moment}  approaches its actual expectation value when $ \delta\rightarrow 0 $. From Fig. \ref{fig:fig1_backlog}, we find that,  the curves for $ \rho_a= 1$ and $ 2 $~Gbps by adopting $ \delta=10^{-2} $ perfectly match with the actual values, i.e., results by $ \delta\rightarrow 0 $. Nevertheless, for $ \rho_{a}=3 $ Gbps and $ \delta=10^{-2} $, the bound overestimates the simulated backlog distribution significantly which suggests that the precision of $ \delta=10^{-2} $ is not sufficient in this case, i.e., more precision is required for highly utilized systems. This suggests a trade-off between computation complexity of the bounds and their tightness. On the other hand, for  practical system operation, a precision of $ \delta=10^{-2} $ is  able to provide acceptable backlog bound for moderate utilization, e.g., $ \rho_{a}\leq 2 $ Gbps. 

\begin{figure}
\centering
\includegraphics[width=3.4in]{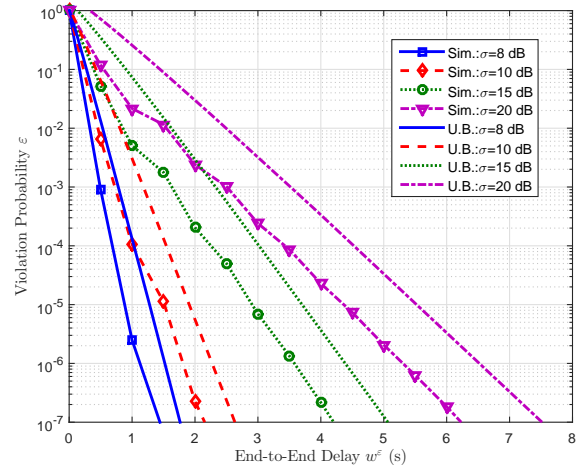}
\caption{Violation probability $ \eps $ v.s.  targeted delay bound $ w^{\eps} $, compared to simulation with $ \rho_a=1 $ Gbps  and $ \kappa=25 $ dB, for different $\sigma$. }
\label{fig:fig2_delay}
\end{figure}

In Fig.~\ref{fig:fig2_delay}, we investigate the delay performance for different shadowing standard deviations $ \sigma $ which is given by the channel characterization in \eqref{eqn:channel gain}. We use $ \delta=10^{-2} $ in the computation of the delay bounds. The figure shows that the analytical delay bounds predict the tail distribution of the simulated delay and are are all asymptotically tight. Furthermore, the delay performance of the system deteriorates as the standard deviation of the shadowing gets larger.  

From the above, we can conclude that the expressions given by  Lemma~\ref{lemma:inverse moment} and Theorem~\ref{theorem:probabilistic bound} provide asymptotically tight bounds for  the probabilistic backlog and delay of $ 60 $~GHz wireless systems. This motivates the use of the computed analytical bounds to analyze and optimize the  performance of realistic and more complex topologies of the $ 60 $ GHz wireless communication networks.



\begin{figure}
\centering
\includegraphics[width=3.4in]{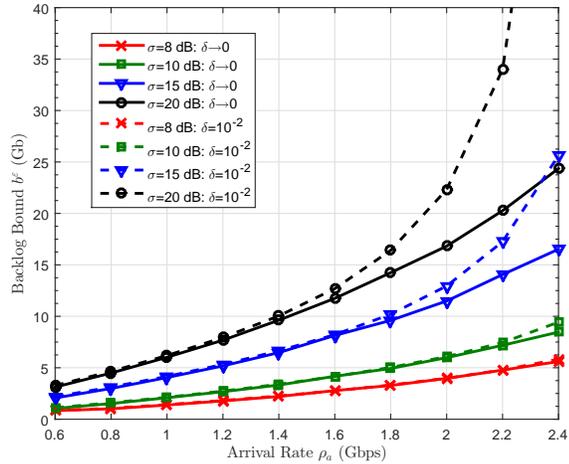}
\caption{Backlog bound $ b^{\eps} $ v.s. arrival rate $ \rho_{a} $, with $ \eps = 10^{-5} $  and $ \kappa=25 $~dB, for different $\sigma$.}
\label{fig:fig3_backlog}
\end{figure}

In Fig.~\ref{fig:fig3_backlog}, with respect to a given violation probability $ \eps=10^{-5} $, we illustrate the probabilistic backlog bound $ b^{\eps} $ against the arrival rate $ \rho_{a} $ for four different channel shadowing distributions. The figure shows that when  the arrival rate $ \rho_{a} $ increases, the precision $ \delta=10^{-2} $ becomes insufficient for  rigorous analysis, as it results in gross overestimation of the backlog bound  compared to the finer precision, i.e., $ \delta\rightarrow 0 $, case.
 This indicates that  smaller values of $ \delta $ are required in order to maintain the precision as the network load becomes higher. 
 We also notice that the shadowing effect, in terms of $ \sigma $, on network performance increases rapidly  as the network load increases. 

\begin{figure}
\centering
\includegraphics[width=3.4in]{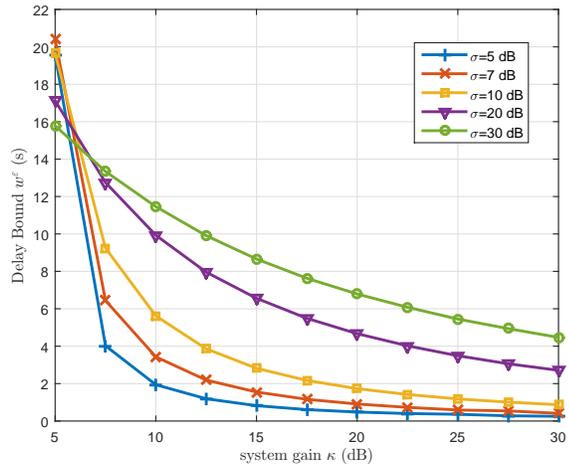}
\caption{Delay bound $ w^{\eps} $ v.s. system gain $ \kappa $,  with $ \delta\rightarrow 0 $, $ \eps = 10^{-3} $ and $ \rho_{a}=1 $~Gbps, for different $ \sigma $.}
\label{fig:fig4_delay}
\end{figure}

According to \eqref{eqn:SNR}, the SNR $ \gamma $ is characterized by the system gain $ \kappa $. Varying $ \kappa $ can be obtained  by manipulating the system configurations, such as antenna gains, transmit power and Tx/Rx distance. 
Fig.~\ref{fig:fig4_delay} plots the probabilistic delay bound against system gain for different shadowing distributions.
In general, by increasing $ \kappa $, the delay bounds remarkably decay, while the benefits gradually diminish. 
For curves regarding different $ \sigma $, an interesting finding from Fig.~\ref{fig:fig4_delay} is that, a smaller value of $ \sigma $ surprisingly gives a higher delay bound $ b^{\eps} $ within the region of lower system gain, i.e., when $ \kappa=5 $ dB, while it also has a faster decay when increasing $ \kappa $. On the other hand, when the system gain is moderate or higher, a larger$\sigma $ will definitely result in more significant performance deterioration in terms of delay bound.


From Fig.~\ref{fig:fig3_backlog} and \ref{fig:fig4_delay} reflects the importance of considering the arrival and fading channel characteristics as well as system gain in the design and implementation of QoS-driven  $ 60 $ GHz wireless networks. 

\section{Conclusions}\label{Sec: VI}
We investigate the probabilistic backlog and delay bounds of the $ 60 $~GHz wireless networks. We provide a MGF-based closed-form expression for the cumulative network service process.  Subsequently,  probabilistic upper bounds on the backlog and delay of the $ 60 $~GHz network are derived using MGF-based network calculus approach. The analytical bounds are validated using extensive  simulations. The obtained results demonstrated  the asymptotic tightness of the analytical performance bounds. We use the obtained analytical bounds to study the impact of several network components, e.g., arrival rate and system gain, on the network performance. The results quantify the potential of these parameters for performance improvement. In   light of the above, we believe that the proposed analytical approach  will have broad applications in designing, optimizing and dimensioning $ 60 $~GHz networks under performance constraints. The approach can be extended to address many of the issues related to the next generation wireless communications in terms of performance improvements and QoS guarantees. For our future work, we are addressing service guarantees in   multi-hop heterogeneous $ 60 $~GHz backhaul networks as well as transmit power management and optimization in such networks. 



\end{document}